\documentclass[pra,aps,superscriptaddress,nofootinbib,twocolumn]
{revtex4-1}
\usepackage{etoolbox}
\apptocmd{\thebibliography}{}{}{}

\usepackage{mathrsfs}
\usepackage{amsfonts}
\usepackage{amssymb}
\usepackage{amsmath}
\usepackage{graphicx}

\usepackage{appendix}

\usepackage[braket]{qcircuit}
\usepackage{algorithm}
\usepackage[noend]{algpseudocode}
\usepackage{algpseudocode}
\usepackage{subfigure,dcolumn}
\usepackage{graphicx}
\usepackage{mathtools}
\usepackage{braket}
\usepackage{subfigure}
\usepackage{epstopdf}
\usepackage[usenames,dvipsnames]{color}
\usepackage[colorlinks=true,citecolor=blue,linkcolor=magenta]{hyperref}
\usepackage{ulem}
\usepackage{lmodern}
\usepackage{amsthm}

\usepackage{listings}
\usepackage{algorithm}  
\usepackage{algpseudocode}  
\usepackage{amsmath}

\usepackage[marginal]{footmisc}
\renewcommand{\thefootnote}
 
\newtheorem{lemma}{Lemma}

\begin{document}

\title{Dominant Eigenstate Isolation through Symmetry-Orbit Decoupling}

\author{Canyu He}
\affiliation{Faculty of Science, Kunming University of Science and Technology, Kunming 650500, China}

\author{Chunhua Zeng}
\affiliation{Faculty of Science, Kunming University of Science and Technology, Kunming 650500, China}

\author{Ruyu Yang}
\email{yangruyu96@gmail.com}
\affiliation{Graduate School of China Academy of Engineering Physics, Beijing 100193, China}

\author{Yongdan Yang}
\email{danyongyang@163.com}
\affiliation{Faculty of Science, Kunming University of Science and Technology, Kunming 650500, China}

\begin{abstract}
We address the problem of extracting a target eigenstate from an approximate pure state. Our approach, symmetry-orbit decoupling, exploits symmetries that preserve the target state while acting differently on the residual. Averaging over the resulting orbit suppresses the target--residual coherence without reducing the target population, thereby exposing a spectral structure that is hidden in the original pure state. When the target is sufficiently separated from the residual modes, it becomes the principal mode of the orbit-averaged state and can be identified through the orbit Gram matrix.  Numerical VQE simulations performed with QuESTlink for six- and eight-qubit Heisenberg and Fermi--Hubbard chains show substantial improvements in ground-state energy estimation. The remaining accuracy is limited by residual components that are indistinguishable from the target under the chosen symmetry.
\end{abstract}


\maketitle

\section{Introduction}
\label{sec:introduction}

Extracting physical properties of eigenstates from approximate quantum
states is a fundamental objective in quantum simulation. A broad class
of quantum algorithms, including variational quantum algorithms and
other hybrid quantum-classical approaches, prepare states that
approximate desired eigenstates with finite quantum resources
\cite{peruzzo2014variational,mcclean2016theory,cerezo2021variational,
tilly2022variational}. However, even with ideal quantum hardware,
finite circuit depth, restricted ansatz expressibility, and incomplete
classical optimization can introduce intrinsic algorithmic residuals
into the prepared state
\cite{sim2019expressibility,du2022efficient,mcclean2018barren, holmes2022connecting}. Distinct from hardware-induced errors~\cite{sun2021realistic}, these residuals arise from the
algorithmic approximation itself and remain coherent within the
prepared state. A central challenge is therefore to separate the
target contribution from the coherent residual structure.

Spectral purification methods provide one route to eigenstate
extraction by amplifying dominant components of a density operator.
Virtual distillation and related approaches suppress subleading
eigenvalues through nonlinear spectral filtering
\cite{koczor2021exponential,huggins2021virtual,koczor2021dominant,
huo2022dual,seif2023shadow,cai2023quantum}. They are particularly
effective when the desired state already appears as a dominant
eigencomponent of a mixed-state density operator. However, a coherent
algorithmic approximation can remain nearly pure. In this case, the
density operator is rank one, and density-matrix powers preserve the
original approximate state rather than isolating the hidden target
component. This motivates the construction of an intermediate ensemble
in which the target contribution can become a distinct spectral mode.

Quantum subspace expansion and its generalized extensions address
algorithmic inaccuracies by enlarging the accessible state space and
solving projected Hamiltonian eigenvalue problems
\cite{mcclean2017hybrid,mcclean2020decoding,yoshioka2022generalized}.
Related nonorthogonal and Krylov-subspace approaches also exploit
structured collections of quantum states and overlap information
\cite{huggins2020nonorthogonal,parrish2019quantum,motta2020determining, cortes2022quantum, parrish2019filter}.
These methods provide systematic improvements through subspace
refinement, but depend on the choice of basis states and the
measurement of projected operators.

A complementary ensemble-based strategy was recently introduced through the distillation of dominant eigenproperties \cite{bako2026exponential}. By generating a temporal ensemble through random real-time evolution, this approach suppresses coherences between energetically distinct eigenstates and extracts dominant eigenstate properties through classical postprocessing. It demonstrates that appropriately constructed ensembles can reveal eigenstate information contained in approximate quantum states. However, the construction relies on controlled Hamiltonian evolution over a time window determined
by the underlying spectral structure.

Here, we introduce symmetry-orbit decoupling, a complementary approach
that constructs such an ensemble using task symmetries rather than time
evolution \cite{bako2026exponential}. Starting from an approximate algorithmic state, we apply symmetry operations that preserve the target eigenstate up to a phase. Averaging over the resulting symmetry orbit suppresses target--residual coherences between inequivalent symmetry representations through group orthogonality \cite{vollbrecht2001entanglement,bartlett2007reference,marvian2014modes}. When the target--residual coherence is removed and the target contribution is spectrally dominant, the target becomes the principal eigenmode of the orbit-averaged state. 

The mechanism of symmetry-orbit decoupling differs from existing
symmetry-based techniques. Symmetry verification and projection remove
or postselect components outside a prescribed symmetry subspace
\cite{bonet2018low,sagastizabal2019experimental,mcardle2019error,lacroix2023symmetry}.
Symmetry expansion instead combines symmetry-related estimators to
reduce observable bias \cite{cai2021quantum}, while circuit-level symmetry protection modifies approximate evolutions to suppress coherent simulation errors \cite{tran2021faster}. In contrast, symmetry-orbit decoupling reshapes the spectral structure and extracts its principal eigenmode. To identify this mode, we employ Gram-matrix-based classical postprocessing. The nonzero spectrum of an ensemble density operator generated from nonorthogonal pure states is encoded in the corresponding Gram matrix \cite{hughston1993complete,jozsa2000distinguishability}. The principal Gram eigenvector determines the dominant orbit mode, while transition
matrix elements provide its observable properties. We demonstrate the framework through
variational ground-state estimation, where the method improves the
recovery of ground-state energies from algorithmically approximate VQE
states. We further analyze the symmetry resources and orbit-size dependence of the method, together with the limitation imposed by symmetry-blind residual components.

This paper is organized as follows.
Section~\uppercase\expandafter{\romannumeral2} develops the theory of
symmetry-orbit decoupling and dominant-mode extraction.
Section~\uppercase\expandafter{\romannumeral3} applies the framework to
variational ground-state estimation and presents numerical results.
Section~\uppercase\expandafter{\romannumeral4} discusses symmetry
resources, orbit-size dependence, and symmetry-blind limitations.
Section~\uppercase\expandafter{\romannumeral5} concludes the paper.

\section{Theory}
\label{sec:theory}

\begin{figure*}[t]
\centering
\includegraphics[width=0.96\textwidth]{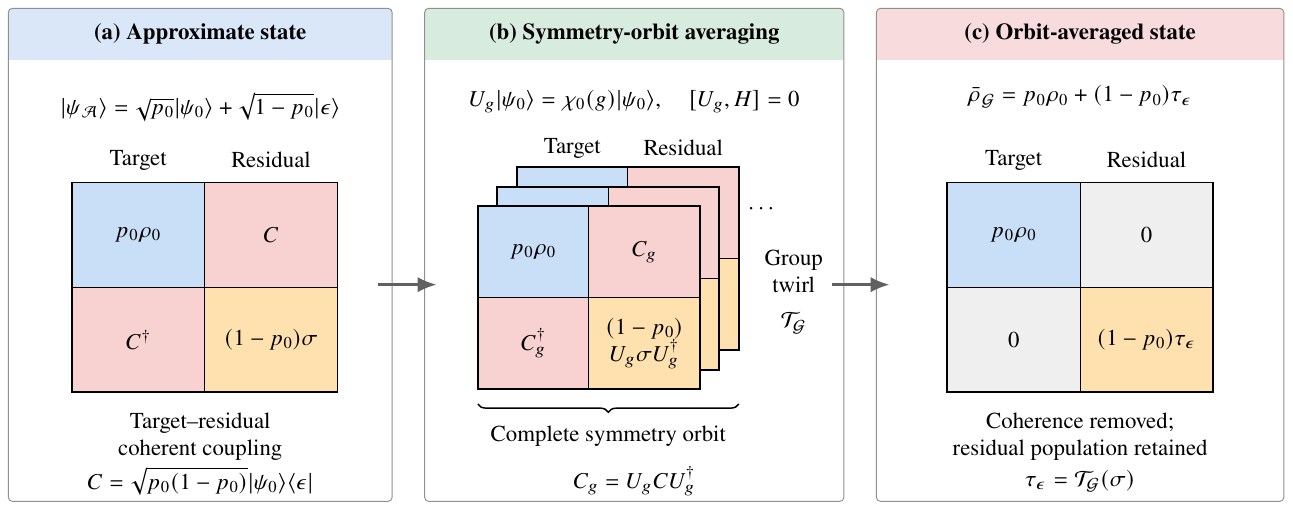}
\caption{
Schematic illustration of symmetry-orbit decoupling. 
(a) The approximate state contains a target population $p_0\rho_0$, a residual population $(1-p_0)\sigma$, and a target--residual coherent coupling $C$. 
(b) Target-preserving symmetry operations $\{U_g\}$ generate the complete symmetry orbit. 
(c) In the exact-decoupling case, group averaging removes the off-diagonal target--residual coherence while retaining the residual population.
}
\label{fig:orbit_decoupling_schematic}
\end{figure*}

In this section, we formulate the problem at the level of the approximate algorithmic state. Throughout this section, we consider ideal quantum hardware so that the deviation from the target state is entirely due to intrinsic algorithmic approximations. 

\subsection{Approximate state and spectral obstruction}

Let $H$ be the problem Hamiltonian and let $\ket{\psi_0}$ be the target eigenstate whose observable properties are to be estimated. An approximate quantum algorithm $\mathcal A$ prepares a normalized pure state $\ket{\psi_{\mathcal A}}$ having target overlap $p_0=|\langle\psi_0|\psi_{\mathcal A}\rangle|^2$. We assume that this overlap is sufficiently large for the target to become spectrally dominant after symmetry-orbit averaging. Without loss of generality, the state can always be decomposed as
\begin{equation}
\ket{\psi_{\mathcal A}}
=
\sqrt{p_0}\,\ket{\psi_0}
+
\sqrt{1-p_0}\,\ket{\epsilon},
\qquad
\langle\psi_0|\epsilon\rangle=0,
\label{eq:state_decomposition}
\end{equation}
where $\ket{\epsilon}$ is the normalized residual component arising from intrinsic algorithmic errors. This decomposition is purely geometrical and does not assume a specific error mechanism: the residual may originate from ansatz limitations, incomplete optimization, or other algorithmic approximations. Since $H$ is Hermitian, its eigenstates form a complete orthonormal basis of the Hilbert space. Therefore, the residual component can be expanded in the eigenbasis as
\begin{equation}
\ket{\epsilon}
=
\sum_{n>0}c_n\ket{\psi_n},
\quad
H\ket{\psi_n}=E_n\ket{\psi_n},
\quad
\sum_{n>0}|c_n|^2=1.
\label{eq:residual_eigenstate_expansion}
\end{equation}
Then, the corresponding density operator of $\ket{\psi_{\mathcal A}}$ is
\begin{equation}
\begin{aligned}
\rho_{\mathcal A}
={}&
p_0\rho_0+(1-p_0)\sigma \\
&+
\sqrt{p_0(1-p_0)}
\sum_{n>0}
\left(
c_n^*\ket{\psi_0}\bra{\psi_n}
+
c_n\ket{\psi_n}\bra{\psi_0}
\right).
\end{aligned}
\label{eq:rho_decomposition}
\end{equation}
where $\rho_0=\ket{\psi_0}\bra{\psi_0}$ and $\sigma =\ket{\epsilon}\bra{\epsilon}$. The first two terms in Eq.~\eqref{eq:rho_decomposition} describe the target and residual populations, whereas the third off-diagonal terms encode their coherent coupling. 

This coupling is the central obstacle to extracting the target contribution. Although $\ket{\psi_0}$ appears in $\ket{\psi_{\mathcal A}}$ with weight $p_0$, it does not correspond to an independent spectral component of $\rho_{\mathcal A}$. Since $\rho_{\mathcal A}$ is a rank-one projector, its only eigenstate with nonzero eigenvalue is the complete approximate state $\ket{\psi_{\mathcal A}}$, up to a global phase. Consequently, any density-matrix-power purification applied directly to the raw state is ineffective, because $\rho_{\mathcal A}^n=\rho_{\mathcal A}$ for every positive integer $n$. This differs fundamentally from the mixed-state setting addressed by virtual distillation \cite{koczor2021exponential,huggins2021virtual}. Our strategy is therefore not to purify the raw state, but first to construct a symmetry-orbit ensemble that removes target--residual coherences.

\subsection{Symmetry orbit and target--residual decoupling}

Let $\mathcal G$ be a finite or compact group represented on the system Hilbert space by unitary operators $U_g$, with $g\in\mathcal G$. The selected symmetry operations are chosen such that they preserve the target eigenstate up to a phase,
\begin{equation}
U_g\ket{\psi_0}
=
\chi_0(g)\ket{\psi_0},
\qquad
|\chi_0(g)|=1,
\label{eq:symmetry_character}
\end{equation}
where $\chi_0(g)$ denotes the phase acquired by the target state. In quantum simulation, such symmetry operations are typically chosen from those satisfying $[U_g,H]=0$. Averaging the approximate state over the resulting symmetry orbit defines the group-twirling channel
\begin{equation}
\mathcal T_{\mathcal G}(X)
=
\int_{\mathcal G}dg\,
U_g X U_g^\dagger,
\qquad
\bar\rho_{\mathcal G}
=
\mathcal T_{\mathcal G}(\rho_{\mathcal A}),
\label{eq:group_twirl}
\end{equation}
where $dg$ denotes the normalized Haar measure. For a finite group, the integral is replaced by the normalized sum $|\mathcal G|^{-1}\sum_{g\in\mathcal G}$. The group twirl acts differently on populations and coherence. In particular, the target population is preserved, while target--residual coherences between inequivalent symmetry representations are removed by group orthogonality. This observation leads to the following decoupling result.

\begin{lemma}[Symmetry-orbit decoupling]
\label{lem:orbit_decoupling}
Suppose that every residual component having nonzero amplitude in
Eq.~\eqref{eq:residual_eigenstate_expansion} transforms in a symmetry
representation different from that of the target. Then the full group
twirl removes the target--residual coherent coupling,
\begin{equation}
\mathcal T_{\mathcal G}
\left(
\ket{\psi_0}\bra{\epsilon}
+
\ket{\epsilon}\bra{\psi_0}
\right)
=
0 .
\label{eq:twirled_coherence}
\end{equation}
Consequently, the orbit-averaged state becomes
\begin{equation}
\bar\rho_{\mathcal G}
=
p_0\rho_0
+
(1-p_0)\tau_\epsilon,
\label{eq:twirled_state}
\end{equation}
where
\begin{equation}
\tau_\epsilon
=
\mathcal T_{\mathcal G}(\sigma).
\label{eq:twirled_residual}
\end{equation}
Since $\tau_\epsilon\ket{\psi_0}=0$, the target state is an exact
eigenstate of $\bar\rho_{\mathcal G}$ with eigenvalue $p_0$.
\end{lemma}

The detailed proof is given in Appendix~\ref{app:coherence_suppression}.
Lemma~\ref{lem:orbit_decoupling} makes precise what is and is not
suppressed by symmetry averaging. Figure~\ref{fig:orbit_decoupling_schematic}
provides a schematic view of this mechanism: the off-diagonal
target--residual coherence is removed by the group twirl, while the
residual population remains in the orbit-averaged state. The twirl
removes target--residual coherences only when the selected symmetry
distinguishes the two components. Residual components transforming in
the same symmetry representation as the target are symmetry blind and
are discussed in Sec.~\ref{subsec:blind_residuals}.

\subsection{Dominant-mode extraction from the symmetry orbit}

Decoupling alone is not sufficient for target-state extraction: the target must also become the leading spectral component of the orbit-averaged state. Under the exact-decoupling condition of Lemma~\ref{lem:orbit_decoupling}, the target contributes the eigenvalue $p_0$, while the residual block contributes eigenvalues bounded by
\begin{equation}
\nu_{\max}
=
\|\tau_\epsilon\|_\infty ,
\label{eq:largest_residual_eigenvalue}
\end{equation}
where $\nu_{\max}$ is the largest eigenvalue of the normalized residual state $\tau_\epsilon$. The target is therefore the unique principal eigenstate if and only if
\begin{equation}
p_0
>
(1-p_0)\nu_{\max}.
\label{eq:dominance_condition}
\end{equation}

The decoupling mechanism does not require the largest available symmetry group. A finite group, or a finite subgroup of a larger symmetry group, is sufficient provided that its action resolves the relevant target--residual coherences. We therefore use the complete orbit of the selected finite group,
\begin{equation}
\ket{\psi_i}
=
U_{g_i}\ket{\psi_{\mathcal A}},
\qquad
i=1,\ldots,N_{\rm orb},
\label{eq:sampled_orbit_states}
\end{equation}
with $N_{\rm orb}=|\mathcal G|$. Its orbit-averaged state is
\begin{equation}
\bar\rho_{N_{\rm orb}}
=
\frac{1}{N_{\rm orb}}
\sum_{i=1}^{N_{\rm orb}}
\ket{\psi_i}\bra{\psi_i},
\label{eq:empirical_orbit_state}
\end{equation}
which coincides with the group average $\bar\rho_{\mathcal G}$ for the selected finite group. The spectral information of $\bar\rho_{N_{\rm orb}}$ can be obtained without constructing the density operator explicitly. Define the orbit Gram matrix
\begin{equation}
S_{ij}
=
\braket{\psi_i|\psi_j}.
\label{eq:orbit_gram_matrix}
\end{equation}
The matrices $\bar\rho_{N_{\rm orb}}$ and $S/N_{\rm orb}$ have identical nonzero spectra (Appendix~\ref{app:gram_spectrum}). Hence, if
\begin{equation}
S\boldsymbol v_0
=
s_0\boldsymbol v_0,
\qquad
s_0\geq s_1\geq\cdots ,
\label{eq:gram_principal_eigenproblem}
\end{equation}
the principal Gram eigenvector directly determines the dominant physical mode,
\begin{equation}
\ket{\phi_0^{(N)}}
=
\frac{1}{\sqrt{s_0}}
\sum_{i=1}^{N_{\rm orb}}
(v_0)_i\ket{\psi_i}.
\label{eq:principal_orbit_mode}
\end{equation}
Under the conditions of Lemma~\ref{lem:orbit_decoupling} and Eq.~\eqref{eq:dominance_condition}, this mode is the target eigenstate up to a global phase. Once the principal mode is identified, an observable $O$ is obtained from the orbit transition matrix
\begin{equation}
O_{ij}^{\rm orb}
=
\bra{\psi_i}O\ket{\psi_j},
\end{equation}
through
\begin{equation}
\langle O\rangle_{\rm orb}
=
\frac{
\boldsymbol v_0^\dagger
O^{\rm orb}
\boldsymbol v_0
}{
\boldsymbol v_0^\dagger
S
\boldsymbol v_0
}.
\end{equation}
Thus, the Gram matrix identifies the dominant orbit mode, while the transition matrix supplies its observable properties.

\section{numerical results}
\label{sec:numerical}

We apply symmetry-orbit decoupling to variational ground-state estimation for two representative lattice models: the Fermi--Hubbard chain and the Heisenberg chain. Both models are studied at six- and eight-qubit sizes. The numerical study tests whether symmetry-orbit decoupling, followed by principal-mode extraction from the orbit Gram matrix, improves ground-state energy estimation from algorithmically approximate VQE states.

All calculations are performed with ideal state-vector simulation, so the deviations from the exact ground states arise from the variational approximation rather than hardware noise. For each system, the VQE parameters are first optimized by minimizing the energy expectation value, after which the resulting state $\ket{\psi_{\mathcal A}}$ is kept fixed for all symmetry choices. For both models, we compare the finite spin-rotation groups $V_4$, $T_{12}$, and $O_{24}$, containing $4$, $12$, and $24$ symmetry branches, respectively, with spatial reflection $\mathcal R=\{I,R\}$ and the combined groups $\mathcal G_s\times\mathcal R$. The complete finite group is used in each calculation. All symmetry operators are verified numerically to be unitary, to commute with the corresponding Hamiltonian, and to preserve the target ray. We quantify the final accuracy by the energy error $\delta E=|E-E_0|$, where $E_0$ is the exact ground-state energy obtained by exact diagonalization.

\begin{figure}[t]
\centering
\includegraphics[width=\columnwidth]{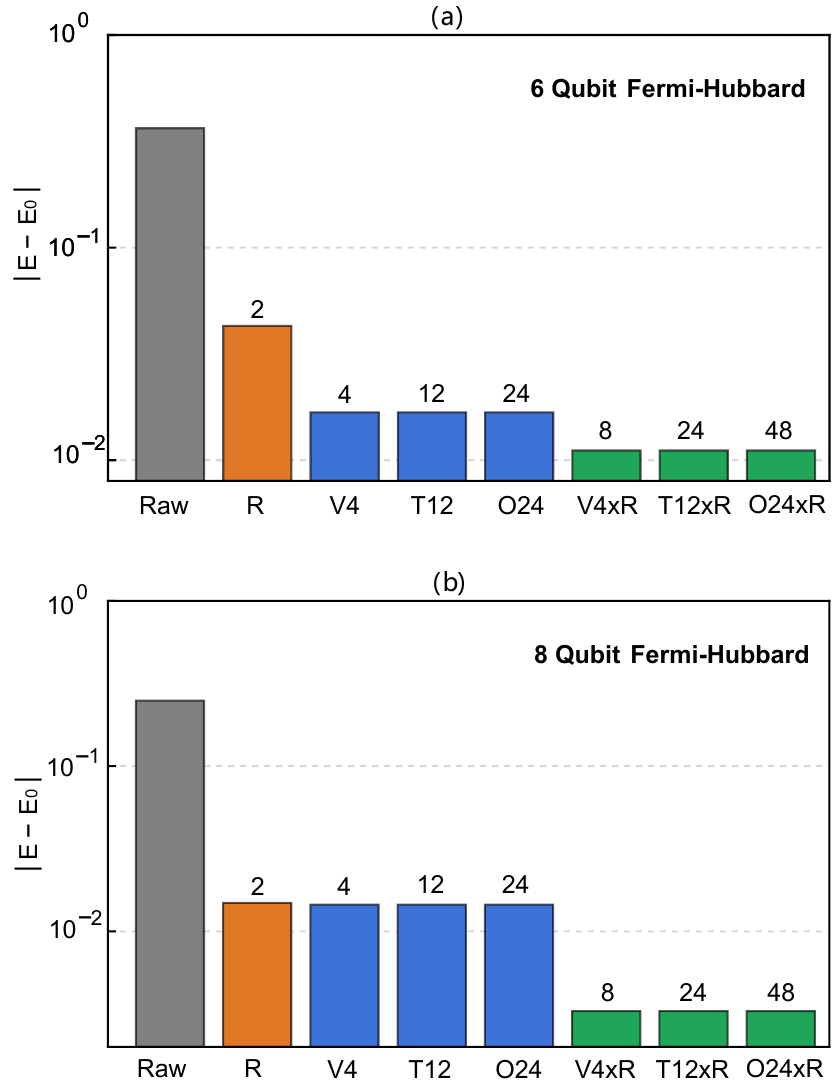}
\caption{
Ground-state energy errors for the Fermi--Hubbard chain with
(a) three sites encoded on six qubits and (b) four sites encoded on
eight qubits. Raw VQE is compared with principal-mode extraction using spin rotations, spatial reflection, and their combined orbits. The numbers above the bars
indicate the orbit size $N_{\rm orb}$ for the corresponding symmetry group.
}
\label{fig:hubbard_results}
\end{figure}

\subsection{Fermi--Hubbard chains}
\label{subsec:hubbard_numerics}

We first consider the Fermi--Hubbard model on open chains of $L=3$ and $4$ sites, corresponding to six and eight qubits. We use the particle--hole symmetric form
\begin{equation}
\begin{aligned}
H_{\rm FH}
={}&
-t\sum_{j=1}^{L-1}\sum_{\sigma=\uparrow,\downarrow}
\left(
c_{j\sigma}^{\dagger}c_{j+1,\sigma}
+
c_{j+1,\sigma}^{\dagger}c_{j\sigma}
\right)
\\
&+
U\sum_{j=1}^{L}
\left(n_{j\uparrow}-\frac12\right)
\left(n_{j\downarrow}-\frac12\right),
\end{aligned}
\label{eq:hubbard_num}
\end{equation}
where \(c_{j\sigma}\) is the annihilation operator for an electron with spin \(\sigma=\uparrow,\downarrow\) on site \(j\), and \(n_{j\sigma}=c_{j\sigma}^{\dagger}c_{j\sigma}\) is the corresponding number operator. The model is mapped to qubits using the Jordan--Wigner transformation. We set $t=2$ and $U=12$ for both lattice sizes. The spin and reflection symmetries are implemented consistently with the fermionic encoding, including the exchange signs associated with spatial reflection.

Figure~\ref{fig:hubbard_results} compares the raw and symmetry-orbit extracted energy errors for both system sizes. For six qubits, the raw VQE error of $0.36413$ is reduced to approximately $4.29\times10^{-2}$ by reflection, to $1.68\times10^{-2}$ by the $T_{12}$ and $O_{24}$ spin orbits, and further to about $1.11\times10^{-2}$ when spin rotations are combined with reflection. The same trend is observed for eight qubits, where the raw error of $0.24925$ is reduced to approximately $1.45\times10^{-2}$ by spin-only extraction, $1.48\times10^{-2}$ by reflection, and $3.29\times10^{-3}$ by the combined symmetry orbit. In both cases, the joint spin--reflection construction gives the most accurate estimate, consistent with the two symmetries resolving different components of the variational residual. The nearly identical results obtained with $T_{12}$ and $O_{24}$ further show that a larger symmetry group does not by itself guarantee a lower extracted energy error.

\begin{figure}[t]
\centering
\includegraphics[width=\columnwidth]{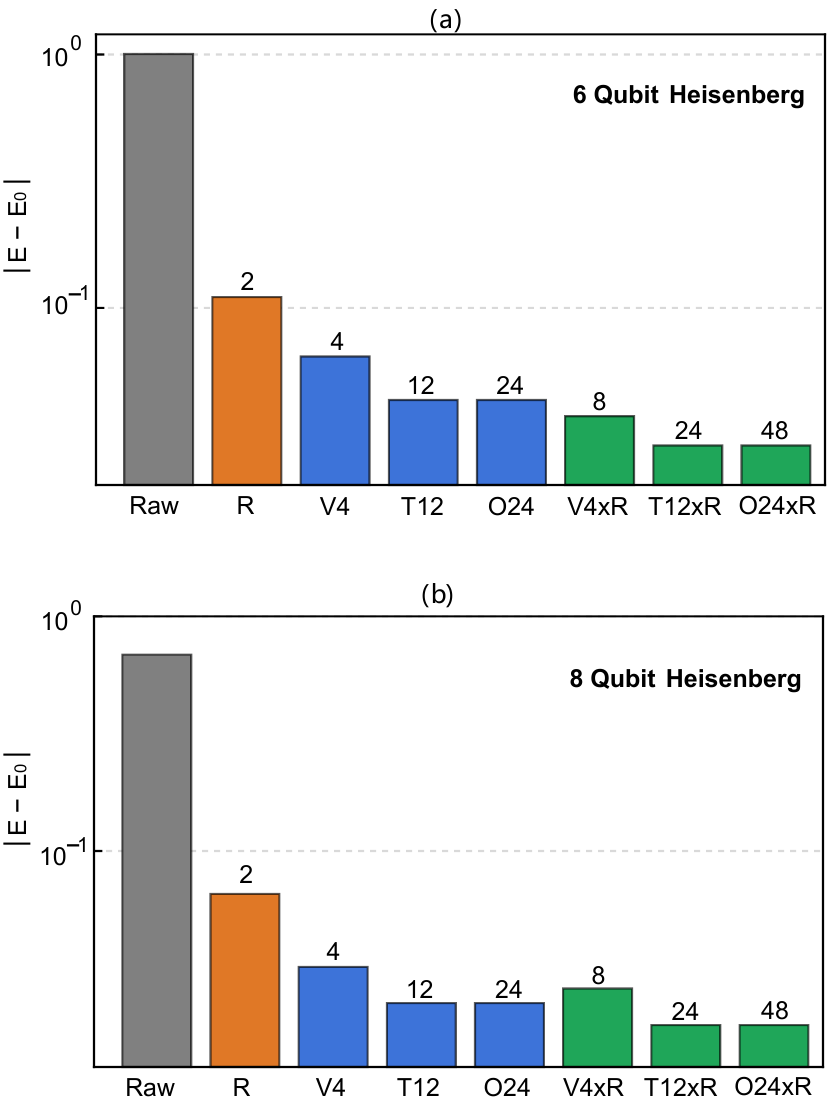}
\caption{
Ground-state energy errors for the Heisenberg chain with
(a) six and (b) eight qubits at $J=1$. Raw VQE is compared with principal-mode extraction using spin rotations, spatial reflection, and their combined orbits. The numbers above the bars indicate the
orbit size $N_{\rm orb}$ for the corresponding symmetry group.
}
\label{fig:heisenberg_results}
\end{figure}

\subsection{Heisenberg chains}
\label{subsec:heisenberg_numerics}

We next consider the open Heisenberg chain
\begin{equation}
H_{\rm H}
=
J\sum_{j=1}^{n_q-1}
\left(
X_jX_{j+1}
+
Y_jY_{j+1}
+
Z_jZ_{j+1}
\right),
\label{eq:heisenberg_num}
\end{equation}
where $X_j$, $Y_j$, and $Z_j$ are the Pauli operators acting on site $j$, $n_q$ is the number of spins (qubits), and $J$ denotes the nearest-neighbor exchange coupling. The spin symmetries are implemented as collective rotations $U_g=u_g^{\otimes n_q}$, while reflection reverses the site ordering about the center of the open chain. For the spin model, the latter is an ordinary permutation of qubits. These two symmetry structures can therefore be tested separately and in combination.

Figure~\ref{fig:heisenberg_results} compares the raw VQE energy errors with those obtained from the principal Gram modes for both chain lengths. For six qubits, the raw error of $1.00$ is reduced to $1.10\times10^{-1}$ by reflection and to $6.43\times10^{-2}$ by the $V_4$ orbit. The larger spin groups provide a further improvement, with both $T_{12}$ and $O_{24}$ reaching an error of approximately $4.33\times10^{-2}$. Combining spin rotations with reflection lowers the error further, with the $T_{12}\times\mathcal R$ and $O_{24}\times\mathcal R$ orbits reaching $2.86\times10^{-2}$. A similar hierarchy is observed for eight qubits: the raw error of $6.87\times10^{-1}$ is reduced to $6.56\times10^{-2}$ by reflection, $3.20\times10^{-2}$ by $V_4$, and $2.24\times10^{-2}$ by both $T_{12}$ and $O_{24}$. The combined $T_{12}\times\mathcal R$ and $O_{24}\times\mathcal R$ orbits give the lowest error, $1.81\times10^{-2}$. Thus, in both system sizes, richer symmetry structure improves the extraction, while the near-identical results for $T_{12}$ and $O_{24}$ show that increasing the orbit size alone does not guarantee additional accuracy.

The results for the two model show that symmetry-orbit extraction improves ground-state energy estimation in both fermionic many-body and spin settings despite the different physical realizations of the underlying symmetries. Because the selected symmetry operations commute with the Hamiltonian, linear orbit averaging alone leaves the energy unchanged; the improvement instead arises from extraction of the principal orbit mode.

\section{Performance and limitations}
\label{sec:performance}

The results of Sec.~\ref{sec:numerical} show that the effectiveness of symmetry-orbit decoupling is determined not simply by the number of symmetry operations, but by how the selected symmetry acts on the residual components of the approximate state. This leads to two practical considerations. First, enlarging the symmetry orbit increases the required resources but does not necessarily improve the extraction accuracy. Second, residual components that cannot be distinguished from the target by the selected symmetry remain coupled to it and therefore set an intrinsic limitation of the method.

\subsection{Symmetry resources and orbit size}
\label{subsec:orbit_resources}

For a finite symmetry group $\mathcal G$, the complete orbit contains
\begin{equation}
N_{\rm orb}=|\mathcal G|
\label{eq:finite_orbit_size}
\end{equation}
symmetry-related branches. The additional quantum cost therefore increases with the number of symmetry operations and with the circuit cost of implementing each $U_g$. For the Heisenberg models considered here, the spin transformations take the form $U_g=u_g^{\otimes n_q}$ and can be implemented by parallel single-qubit rotations, so each individual symmetry operation adds only a shallow circuit layer compared with the VQE state preparation. The measurement effort required for the Gram and transition matrices also increases with the orbit size, since more symmetry branches generate more matrix elements to be evaluated.

A larger orbit, however, does not necessarily produce a more accurate extracted mode. In the six-qubit Heisenberg model, increasing the spin group from $T_{12}$ to $O_{24}$ doubles $N_{\rm orb}$ from $12$ to $24$, whereas the two cases reach an error of approximately $4.33\times10^{-2}$. Similar saturation is observed in the Fermi--Hubbard calculations. These results show that $N_{\rm orb}$ is primarily a resource parameter rather than a universal measure of accuracy. Additional symmetry operations are useful only when they provide new resolution of residual components that remain unresolved by the smaller symmetry set.

\subsection{Symmetry-blind residuals and limitations}
\label{subsec:blind_residuals}

The more fundamental limitation of the method is that the selected symmetry may not distinguish every residual component from the target. Residual components that transform differently from the target under the group are \emph{symmetry-visible}; group averaging removes their coherence with the target, as described in Sec.~\ref{sec:theory}. By contrast, residual components transforming in the same symmetry representation as the target are \emph{symmetry-blind}. Because the selected symmetry cannot distinguish these components from the target, their coherence is not removed by the group average and they cannot be decoupled by that symmetry alone.

To separate these two contributions, let $\Pi_0$ denote the projector onto the full symmetry subspace transforming according to the same representation as the target. Its role is simply to select from the residual the part that is indistinguishable from the target under the chosen symmetry. We therefore write
\begin{equation}
\ket{\epsilon}
=
\ket{\epsilon_{\rm b}}
+
\ket{\epsilon_{\rm v}},
\label{eq:blind_visible_decomposition}
\end{equation}
with
\begin{equation}
\ket{\epsilon_{\rm b}}
=
\Pi_0\ket{\epsilon},
\qquad
\ket{\epsilon_{\rm v}}
=
(I-\Pi_0)\ket{\epsilon}.
\label{eq:blind_visible_components}
\end{equation}
Here $\ket{\epsilon_{\rm b}}$ and $\ket{\epsilon_{\rm v}}$ denote the symmetry-blind and symmetry-visible parts of the residual, respectively. Their relative importance is characterized by the blind weight
\begin{equation}
\eta_0
=
\langle\epsilon_{\rm b}|\epsilon_{\rm b}\rangle
=
\bra{\epsilon}\Pi_0\ket{\epsilon},
\qquad
0\leq\eta_0\leq1 .
\label{eq:blind_weight}
\end{equation}
The action of the group average on the two components is then immediate. Since the visible part transforms differently from the target, group orthogonality gives
\begin{equation}
\mathcal T_{\mathcal G}
\left(
\ket{\psi_0}\bra{\epsilon_{\rm v}}
\right)
=0 .
\label{eq:visible_coherence}
\end{equation}
The blind part, however, transforms in the same representation as the target and therefore survives,
\begin{equation}
\mathcal T_{\mathcal G}
\left(
\ket{\psi_0}\bra{\epsilon_{\rm b}}
\right)
=
\ket{\psi_0}\bra{\epsilon_{\rm b}} .
\label{eq:blind_coherence_survives}
\end{equation}
Combining Eqs.~\eqref{eq:blind_visible_decomposition}--\eqref{eq:blind_coherence_survives}, the target--residual coherence remaining after the group average is
\begin{equation}
\mathcal T_{\mathcal G}
\left(
\ket{\psi_0}\bra{\epsilon}
\right)
=
\ket{\psi_0}\bra{\epsilon_{\rm b}} .
\label{eq:blind_coherence_twirl}
\end{equation}
Equation~\eqref{eq:blind_coherence_twirl} shows that the remaining limitation is controlled by the amount of residual weight that is symmetry blind. The quantity $\eta_0$ therefore provides a direct measure of how much of the residual cannot be distinguished from the target by the selected symmetry. A small $\eta_0$ is favorable for target--residual decoupling, but it is not by itself sufficient for accurate extraction: the target contribution must also be spectrally separated from the residual modes.

\begin{figure}[t]
\centering
\includegraphics[width=\columnwidth]{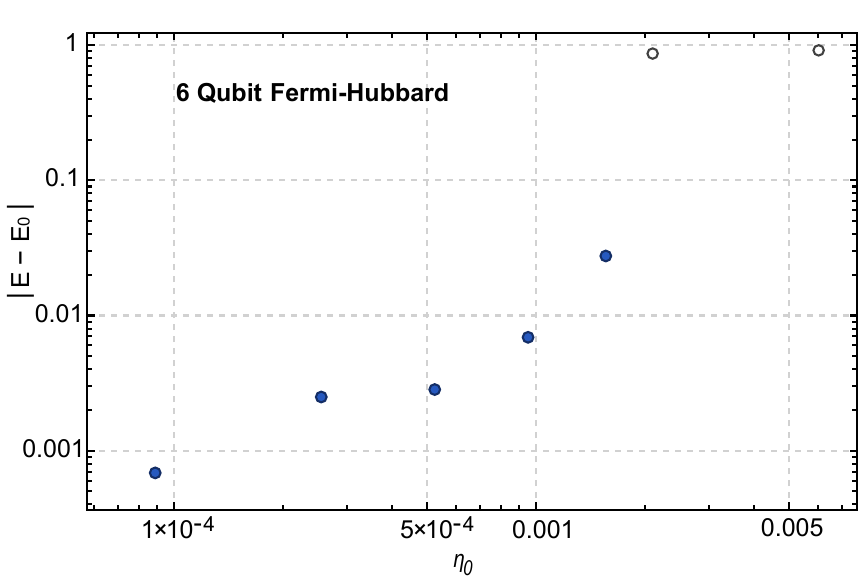}
\caption{
Gram-extracted energy error $|E-E_0|$ versus symmetry-blind residual weight $\eta_0$ for selected VQE states of the six-qubit Fermi--Hubbard model using the complete $O_{24}\times\mathcal R$ orbit. Open circles denote states that do not satisfy the spectral-dominance criterion $p_0>(1-p_0)\nu_{\max}$, while filled circles denote states that satisfy it. Once this criterion is met, smaller $\eta_0$ is generally associated with lower extraction error. Both axes are logarithmic.
}
\label{fig:blind_limit}
\end{figure}

We illustrate this behavior in Fig.~\ref{fig:blind_limit} using selected VQE outputs of the six-qubit Fermi--Hubbard model with the complete $O_{24}\times\mathcal R$ orbit. The open and filled circles distinguish states that do not satisfy and satisfy, respectively, the sufficient condition $p_0>(1-p_0)\nu_{\max}$. The results show that a small symmetry-blind weight alone does not guarantee accurate extraction when spectral dominance is not ensured. Once this condition is satisfied, smaller $\eta_0$ is generally associated with lower Gram-extracted energy error, indicating that symmetry-blind residuals limit the attainable accuracy when the target-associated mode is dominant.

\section{Conclusion}
\label{sec:conclusion}

We have introduced symmetry-orbit decoupling for extracting target eigenstate properties from algorithmically approximate quantum states. By averaging over target-preserving symmetry orbits, the method suppresses target--residual coherences between inequivalent symmetry representations. Under the exact-decoupling and spectral-dominance conditions, the target becomes the principal mode of the orbit-averaged state and can be identified from the orbit Gram matrix. Numerical VQE simulations of six- and eight-qubit Heisenberg and Fermi--Hubbard chains show that principal-mode extraction can substantially improve ground-state energy estimates. The results also show that increasing the orbit size alone does not guarantee higher accuracy; performance depends on how the selected symmetry acts on the residual components.

Residuals transforming in the same symmetry representation as the target remain symmetry blind and cannot be removed by that symmetry alone. The attainable accuracy therefore depends on both spectral separation and the symmetry structure of the residual. Symmetry-orbit decoupling thus provides a distinct route to dominant eigenstate extraction based on symmetry-structured ensemble construction and Gram-matrix postprocessing.

\section*{Data Availability}

The numerical data and source code supporting the findings of this
study are available from the corresponding author upon reasonable
request.

\begin{acknowledgments}
This work is supported by Yunnan Fundamental Research Projects (grant NO.202401BE070001-018) and National Natural Science Foundation of China (Grant No.12565002).
\end{acknowledgments}

\bibliography{ref.bib}
\onecolumngrid
\appendix
\section{Proof of symmetry-orbit decoupling}
\label{app:coherence_suppression}

We prove Lemma~\ref{lem:orbit_decoupling} using the representation
structure of the symmetry group. Since
\begin{equation}
U_g\ket{\psi_0}
=
\chi_0(g)\ket{\psi_0},
\end{equation}
the target state spans a one-dimensional representation of
$\mathcal G$. Under the assumption of
Lemma~\ref{lem:orbit_decoupling}, the residual can be decomposed as
\begin{equation}
\ket{\epsilon}
=
\sum_{\lambda\not\simeq\lambda_0}
\ket{\epsilon_\lambda},
\label{eq:app_residual_rep_decomposition}
\end{equation}
where $\lambda_0$ denotes the representation of the target and each
$\ket{\epsilon_\lambda}$ belongs to a representation $\lambda$
inequivalent to $\lambda_0$. Consider the group-averaged target--residual coherence
\begin{equation}
\mathcal T_{\mathcal G}
\left(
\ket{\psi_0}\bra{\epsilon_\lambda}
\right)
=
\int_{\mathcal G}dg\,
U_g\ket{\psi_0}\bra{\epsilon_\lambda}U_g^\dagger .
\label{eq:app_component_twirl}
\end{equation}
Since the two representations are inequivalent, Schur's lemma gives
\begin{equation}
\mathcal T_{\mathcal G}
\left(
\ket{\psi_0}\bra{\epsilon_\lambda}
\right)
=
0,
\qquad
\lambda\not\simeq\lambda_0 .
\label{eq:app_component_vanishes}
\end{equation}
For one-dimensional representations, this reduces to the usual
orthogonality of distinct group characters. Using Eq.~\eqref{eq:app_residual_rep_decomposition}, we therefore obtain
\begin{equation}
\mathcal T_{\mathcal G}
\left(
\ket{\psi_0}\bra{\epsilon}
\right)
=
0.
\label{eq:app_twirled_target_residual}
\end{equation}
Taking the Hermitian conjugate gives
\begin{equation}
\mathcal T_{\mathcal G}
\left(
\ket{\epsilon}\bra{\psi_0}
\right)
=
0,
\end{equation}
and hence
\begin{equation}
\mathcal T_{\mathcal G}
\left(
\ket{\psi_0}\bra{\epsilon}
+
\ket{\epsilon}\bra{\psi_0}
\right)
=
0.
\label{eq:app_coherence_vanishes}
\end{equation}
Applying the twirl to Eq.~\eqref{eq:rho_decomposition} then yields
\begin{equation}
\bar\rho_{\mathcal G}
=
p_0\rho_0
+
(1-p_0)\tau_\epsilon,
\qquad
\tau_\epsilon
=
\mathcal T_{\mathcal G}(\sigma).
\label{eq:app_twirled_state}
\end{equation}
Because the target ray is preserved by every $U_g$ and
$\langle\psi_0|\epsilon\rangle=0$, each symmetry-transformed residual
remains orthogonal to $\ket{\psi_0}$. Therefore,
\begin{equation}
\tau_\epsilon\ket{\psi_0}=0.
\label{eq:app_residual_orthogonality}
\end{equation}
Consequently,
\begin{equation}
\bar\rho_{\mathcal G}\ket{\psi_0}
=
p_0\ket{\psi_0},
\label{eq:app_target_eigenvalue}
\end{equation}
which proves Lemma~\ref{lem:orbit_decoupling}.

\section{Gram-spectrum equivalence}
\label{app:gram_spectrum}

We prove the spectral equivalence between the orbit Gram matrix and
the empirical orbit state. Define
\begin{equation}
A
=
\sum_{i=1}^{N_{\rm orb}}
\ket{\psi_i}\bra{i},
\label{eq:appendix_A_operator}
\end{equation}
where $\{\ket{i}\}$ is an orthonormal basis of an auxiliary
$N_{\rm orb}$-dimensional space. Then,
\begin{equation}
S=A^\dagger A,
\qquad
\bar\rho_{N_{\rm orb}}
=
\frac{1}{N_{\rm orb}}AA^\dagger .
\label{eq:appendix_A_relation}
\end{equation}
Suppose that
\begin{equation}
S\boldsymbol v_\alpha
=
s_\alpha\boldsymbol v_\alpha,
\qquad
s_\alpha>0 .
\label{eq:appendix_gram_eigen}
\end{equation}
Applying $A$ gives
\begin{equation}
AA^\dagger
\left(
A\boldsymbol v_\alpha
\right)
=
s_\alpha
\left(
A\boldsymbol v_\alpha
\right).
\label{eq:appendix_mapping}
\end{equation}
Thus, $A\boldsymbol v_\alpha$ is an eigenvector of
$AA^\dagger$ with eigenvalue $s_\alpha$. Since
$\bar\rho_{N_{\rm orb}}=AA^\dagger/N_{\rm orb}$, the corresponding
eigenvalue of the empirical orbit state is $s_\alpha/N_{\rm orb}$. Conversely, the nonzero eigenvalues of $AA^\dagger$ and
$A^\dagger A$ coincide. Hence,
\begin{equation}
\operatorname{spec}_{\neq0}
\left(
\bar\rho_{N_{\rm orb}}
\right)
=
\operatorname{spec}_{\neq0}
\left(
\frac{S}{N_{\rm orb}}
\right).
\label{eq:appendix_spectral_equivalence}
\end{equation}
This establishes the Gram-spectrum equivalence used in
Sec.~\ref{sec:theory}.

\end{document}